# Exact solutions of a spatially-dependent mass Dirac equation for Coulomb field plus tensor interaction via Laplace transformation method

M. Eshghi [1,*], S.M. Ikhdair [2]

[1] *Physics Department, Imam Hossein Comprehensive University, Tehran, Iran*
[2] *Physics Department, Near East University, 922022, Nicosia, North Cyprus, Turkey*
*and*
*Physics Department, Faculty of Science, An-Najah National University, Nablus, West Bank, Palestine*
Email: sikhdair@neu.edu.tr

## Abstract

The spatially-dependent mass Dirac equation is solved exactly for attractive scalar and repulsive vector Coulomb potentials including a tensor interaction potential under the spin and pseudospin (p-spin) symmetric limits by using the Laplace transformation method (LTM). Closed forms of the energy eigenvalue equation and wave functions are obtained for arbitrary spin-orbit quantum number $\kappa$. Some numerical results are given too. The effect of the tensor interaction on the bound states is presented. It is shown that the tensor interaction removes the degeneracy between two states in the spin doublets. We also investigate the effects of the spatially-dependent mass on the bound states under the conditions of the spin symmetric limit and in the absence of tensor interaction ($T=0$).

## 1. Introduction

---

[*] Corresponding author: *eshgi54@gmail.com*. Fax: +98 21 77104938.

Dirac equation has become the most appealing relativistic wave equations for spin-1/2 particles. However, solving such a wave equation is still a very challenging problem even if it has been derived more than 80 years ago and has been utilized profusely. It is always useful to investigate the relativistic effects [1-4]. For example, in the relativistic treatment of nuclear phenomena the Dirac equation is used to describe the behavior of the nuclei in nucleus and also in solving many problems of high-energy physics and chemistry. For this reason, it has been used extensively to study the relativistic heavy ion collisions, heavy ion spectroscopy and more recently in laser–matter interaction (for a review, see [5] and references therein) and condensed matter physics [6].

On the other hand, systems with position-dependent mass (PDM) have been found to be very useful in studying the physical properties of various microstructures [7-14]. Recently, there has been increased interest in searching for analytical solutions of the Dirac equation with PDM and with constant mass under the spin and p-spin symmetries [15-32].

Here, we shall attempt to solve the Dirac equation by using the Laplace transform method (LTM). The LTM is an integral transform and is comprehensively useful in physics and engineering [33] and recently used by many authors to solve the Schrodinger equation for different potential forms [34-38]. This method could be a nearly new formalism in the literature and serve as a powerful algebraic treatment for solving the second-order differential equations. As a result, the LTM describes a simple way for solving of radial and one-dimensional differential equations. The other advantage of this method is that a second-order equation can be converted into more simpler form whose solutions may be obtained easily [34]. In this letter, we obtain solution of the Dirac equation both PDM and tensor interaction for attractive scalar and repulsive vector Coulomb potential under the spin and p-spin symmetry limits. We give some numerical results.

## 2. Review to Dirac Equation including Tensor Coupling

The Dirac equation which describes a nucleon in repulsive vector $V(r)$ and attractive scalar $S(r)$ and a tensor $U(r)$ potentials is written as

$$[\vec{\alpha}.\vec{p} + \beta(M(r)c^2 + S(r)) - i\beta\vec{\alpha}.\hat{r}U(r)]\psi(\vec{r}) = [E - V(r)]\psi(\vec{r}), \tag{1}$$

where $M(r)$ is the effective mass of the fermionic particle, $E$ is the relativistic energy of the system, $\vec{p}=-i\hbar\vec{\nabla}$ is the three-dimensional momentum operator. $\vec{\alpha}$ and $\beta$ are the $4\times 4$ Dirac matrices give as

$$\vec{\alpha}=\begin{pmatrix} 0 & \vec{\sigma} \\ \vec{\sigma} & 0 \end{pmatrix}, \quad \beta=\begin{pmatrix} I & 0 \\ 0 & -I \end{pmatrix}, \tag{2}$$

where $I$ is $2\times 2$ unitary matrix and $\vec{\sigma}$ are three-vector spin matrices

$$\sigma_1=\begin{pmatrix} 0 & 1 \\ 1 & 0 \end{pmatrix}, \quad \sigma_2=\begin{pmatrix} 0 & -i \\ i & 0 \end{pmatrix}, \quad \sigma_3=\begin{pmatrix} 1 & 0 \\ 0 & -1 \end{pmatrix}. \tag{3}$$

The total angular momentum operator $\vec{J}$ and spin-orbit $K=(\vec{\sigma}.\vec{L}+1)$, where $\vec{L}$ is orbital angular momentum, of the spherical nucleons commute with the Dirac Hamiltonian. The eigenvalues of spin-orbit coupling operator are $\kappa=(j+1/2)>0$ and $\kappa=-(j+1/2)<0$ for unaligned spin $j=l-1/2$ and the aligned spin $j=l+1/2$, respectively. ($H^2,K,J^2,J_z$) can be taken as the complete set of the conservative quantities. Thus, the spinor wave functions can be classified according to their angular momentum $j$, spin-orbit quantum number $\kappa$ and the radial quantum number $n$ can be written as follows

$$\psi_{n\kappa}(\vec{r})=\begin{pmatrix} f_{n\kappa}(\vec{r}) \\ g_{n\kappa}(\vec{r}) \end{pmatrix}=\frac{1}{r}\begin{pmatrix} F_{n\kappa}(r)Y_{jm}^{l}(\theta,\varphi) \\ iG_{n\kappa}(r)Y_{jm}^{\tilde{l}}(\theta,\varphi) \end{pmatrix}, \tag{4}$$

where $f_{n\kappa}(\vec{r})$ is the upper (large) component and $g_{n\kappa}(\vec{r})$ is the lower (small) component of the Dirac spinors. $Y_{jm}^{l}(\theta,\varphi)$ and $Y_{jm}^{\tilde{l}}(\theta,\varphi)$ are spin and p-spin spherical harmonics, respectively, and $m$ is the projection of the angular momentum on the $z$-axis. Substituting Eq. (4) into Eq. (1) and using the following relations [39] as

$$(\vec{\sigma}.\vec{A})(\vec{\sigma}.\vec{B})=\vec{A}.\vec{B}+i\vec{\sigma}.(\vec{A}\times\vec{B}), \tag{5a}$$

$$(\vec{\sigma}.\vec{P})=\vec{\sigma}.\hat{r}\left(\hat{r}.\vec{P}+i\frac{\vec{\sigma}.\vec{L}}{r}\right), \tag{5b}$$

and with the following properties

$$\vec{\sigma}.\vec{L}\begin{cases} Y_{jm}^{\tilde{l}}(\theta,\phi) \\ Y_{jm}^{l}(\theta,\phi) \end{cases}=\begin{cases} (\kappa-1)Y_{jm}^{\tilde{l}}(\theta,\phi), \\ -(\kappa-1)Y_{jm}^{l}(\theta,\phi), \end{cases} \tag{6a}$$

$$\vec{\sigma}\cdot\hat{r}\begin{cases}Y^{\tilde{l}}_{jm}(\theta,\phi)\\ Y^{l}_{jm}(\theta,\phi)\end{cases}=\begin{cases}-Y^{l}_{jm}(\theta,\phi),\\ -Y^{\tilde{l}}_{jm}(\theta,\phi),\end{cases} \quad (6b)$$

one obtains two coupled differential equations for upper and lower radial wave functions $F_{n\kappa}(r)$ and $G_{n\kappa}(r)$ as

$$\left(\frac{d}{dr}+\frac{\kappa}{r}-U(r)\right)F_{n\kappa}(r)=\left(M(r)c^2+E_{n\kappa}-\Delta(r)\right)G_{n\kappa}(r), \quad (7a)$$

$$\left(\frac{d}{dr}-\frac{\kappa}{r}+U(r)\right)G_{n\kappa}(r)=\left(M(r)c^2-E_{n\kappa}+\Sigma(r)\right)F_{n\kappa}(r). \quad (7b)$$

where

$$\Delta(r)=V(r)-S(r), \quad (8a)$$

$$\Sigma(r)=V(r)+S(r). \quad (8b)$$

Eliminating $F_{n\kappa}(r)$ and $G_{n\kappa}(r)$ from Eqs. (7), we finally obtain the following two Schrödinger-like differential equations for the upper and lower radial spinor components, respectively:

$$\begin{aligned}&\left[\frac{d^2}{dr^2}-\frac{\kappa(\kappa+1)}{r^2}+\frac{2\kappa}{r}U(r)-\frac{dU(r)}{dr}-U^2(r)\right]F_{n\kappa}(r)\\ &-\frac{\frac{dM(r)}{dr}-\frac{d\Delta(r)}{dr}}{M(r)+E_{n\kappa}-\Delta(r)}\left(\frac{d}{dr}+\frac{\kappa}{r}-U(r)\right)F_{n\kappa}(r)\\ &=\frac{1}{(\hbar c)^2}\left[\left(M(r)c^2+E_{n\kappa}-\Delta(r)\right)\left(M(r)c^2-E_{n\kappa}+\Sigma(r)\right)\right]F_{n\kappa}(r),\end{aligned} \quad (9a)$$

and

$$\begin{aligned}&\left[\frac{d^2}{dr^2}-\frac{\kappa(\kappa-1)}{r^2}+\frac{2\kappa}{r}U(r)+\frac{dU(r)}{dr}-U^2(r)\right]G_{n\kappa}(r)\\ &-\frac{\frac{dM(r)}{dr}+\frac{d\Sigma(r)}{dr}}{M(r)-E_{n\kappa}+\Sigma(r)}\left(\frac{d}{dr}-\frac{\kappa}{r}+U(r)\right)G_{n\kappa}(r)\\ &=\frac{1}{(\hbar c)^2}\left[\left(M(r)c^2+E_{n\kappa}-\Delta(r)\right)\left(M(r)c^2-E_{n\kappa}+\Sigma(r)\right)\right]G_{n\kappa}(r),\end{aligned} \quad (9b)$$

where $\kappa(\kappa-1)=\tilde{l}(\tilde{l}+1)$ and $\kappa(\kappa+1)=l(l+1)$. These radial wave functions are required to satisfy the necessary boundary conditions. The spin-orbit quantum number $\kappa$ is related to the quantum numbers for spin symmetry $l$ and p-spin symmetry $\tilde{l}$ as

$$\kappa = \begin{cases} -(l+1) = -(j+1/2), & (s_{1/2}, p_{3/2}, etc.) \quad j = l + \frac{1}{2}, \text{ aligned spin}(\kappa < 0), \\ l = j+1/2, & (p_{1/2}, d_{3/2}, etc.) \quad j = l - \frac{1}{2}, \text{ unaligned spin}(\kappa > 0), \end{cases} \quad (10)$$

and the quasi-degenerate doublet structure can be expressed in terms of a p-spin angular momentum $\tilde{s} = 1/2$ and pseudo-orbital angular momentum $\tilde{l}$, which is defined as

$$\kappa = \begin{cases} -\tilde{l} = -(j+1/2); & (s_{1/2}, p_{3/2}, etc.) \quad j = \tilde{l} - \frac{1}{2}, \text{ aligned p-spin}(\kappa < 0), \\ \tilde{l}+1 = j+1/2; & (d_{3/2}, f_{5/2}, etc.) \quad j = \tilde{l} + \frac{1}{2}, \text{ unaligned p-spin}(\kappa > 0), \end{cases} \quad (11)$$

where $\kappa = \pm 1, \pm 2, \dots$. For example, $(1s_{1/2}, 0d_{3/2})$ and $(1p_{3/2}, 0f_{5/2})$ can be considered as p-spin doublets. In the next section, we will consider the spin and p-spin symmetry cases.

## 2. Relativistic Bound State Solutions
## 2.1. Spin symmetry case

Equation (9a) can not be solved analytically because of the last term $\frac{dM(r)/dr - d\Delta(r)/dr}{M(r) + E_{n\kappa} - \Delta(r)} \left( \frac{d}{dr} + \frac{\kappa}{r} - U(r) \right)$. Therefore, in solving the mathematical relation $dM(r)c^2/dr = d\Delta(r)/dr = dV(r)/dr$ [40, 41], we can then exactly solve Eq. (9a). In this stage, we take the vector potential in the form of an attractive Coulomb-like field [18] as

$$\Sigma(r) = V(r) = -\frac{\hbar c q_v}{r}, \quad q_v = q, \quad r \neq 0, \quad (12)$$

where $q_v$ is being a vector dimensionless real parameter coupling constant and $\hbar c$ is being a constant with $J.fm$ dimension. Also, it is convenient to take the mass function [18] as

$$M(r) = m_0 + \frac{m_1}{r}, \quad m_1 = m_0 \lambda_0 b, \quad \lambda_0 = \frac{\hbar}{m_0 c}, \quad (13)$$

where $m_0$ and $m_1$ stand for the rest mass of the fermionic particle and the perturbed mass, respectively. Further, $b$ is the dimensionless real constant to be set to zero for the constant mass case and $\lambda_0$ is the Compton-like wavelength in *fm* units. Further, the tensor interaction takes the simple form:

$$U(r) = -\frac{T}{r}, \qquad T = \frac{Z_a Z_b e^2}{4\pi\varepsilon_0}, \qquad r \geq R_c, \tag{14}$$

where $R_c$ is the coulomb radius, $Z_a$ and $Z_b$ stand for the charges of the projectile particle *a* and the target nucleus *b*, respectively.

Substituting Eqs. (12)-(14) into (9a) considering the spin symmetry case where $\Delta(r) = C_s = \text{const.}$, i.e., $(d\Delta(r)/dr = 0)$ [42, 43]. Thus, the equation obtained for the upper component of the Dirac spinor $F_{n\kappa}(r)$ becomes

$$\left\{ \frac{d^2}{dr^2} - \frac{(\kappa+T)(\kappa+T+1) + b(b-q)}{r^2} \right.$$
$$+ \frac{1}{\hbar c} \frac{\left[ q\left(m_0 c^2 + E_{n\kappa} - C_s\right) + b\left(C_s - 2m_0 c^2\right) \right]}{r} \tag{15}$$
$$\left. - \frac{1}{(\hbar c)^2}(m_0 c^2 - E_{n\kappa})(m_0 c^2 + E_{n\kappa} - C_s) \right\} F_{n\kappa}(r) = 0.$$

Further, defining the new parameters

$$\lambda^2 = (\kappa+T)(\kappa+T+1) + b(b-q),$$
$$\delta^2 = -\frac{1}{\hbar c}\left[ q\left(m_0 c^2 + E_{n\kappa} - C_s\right) + b\left(C_s - 2m_0 c^2\right) \right], \tag{16}$$
$$\varepsilon^2 = \frac{1}{(\hbar c)^2}(m_0 c^2 - E_{n\kappa})(m_0 c^2 + E_{n\kappa} - C_s),$$

and introducing $F_{n\kappa}(r) = r^{1/2}\varphi(r)$, then Eq. (15) turns into the form

$$\left[ r^2 \frac{d^2}{dr^2} + r\frac{d}{dr} - r^2\left( \frac{v^2}{r^2} + \frac{\delta^2}{r} + \varepsilon^2 \right) \right]\varphi(r) = 0, \tag{17a}$$

$$v^2 = \left(\kappa + T + \frac{1}{2}\right)^2 + b(b-q) \tag{17b}$$

Setting $\varphi(r) = r^\alpha f(r)$ with $\alpha$ is a constant and then inserting into Eq. (17a), we have

$$\left[ r^2 \frac{d^2}{dr^2} + (2\alpha+1)r\frac{d}{dr} - \left(v^2 + \delta^2 r + \varepsilon^2 r^2 - \alpha^2\right) \right] f(r) = 0. \tag{18}$$

Now, to obtain a finite wave function when $r \to \infty$, if we take $\alpha = -v$ in equation (18) then it becomes

$$\left[ r \frac{d^2}{dr^2} - (2\nu - 1)\frac{d}{dr} - \left(\delta^2 + \varepsilon^2 r\right) \right] f(r) = 0. \tag{19}$$

The LTM [44,45]

$$L\{g(\rho)\} = f(t) = \int_0^\infty e^{-t\rho} g(\rho) d\rho, \tag{20}$$

leads to an equation

$$\left(t^2 - \varepsilon^2\right)\frac{df(t)}{dt} + \left[(2\nu + 1)t + \delta^2\right] f(t) = 0. \tag{21}$$

Equation (21) is a first-order differential equation and therefore we may directly make use of the integral to get the expression

$$f(t) = N(t+\varepsilon)^{-(2\nu+1)} \left(\frac{t-\varepsilon}{t+\varepsilon}\right)^{-\frac{\left[\delta^2 + (2\nu+1)\varepsilon\right]}{2\varepsilon}}, \tag{22}$$

where $N$ is a constant. Noting that $\left(\frac{t-\varepsilon}{t+\varepsilon}\right)^{-\frac{\delta^2 + (2\nu+1)\varepsilon}{2\varepsilon}}$ is a multi-valued function and the wave functions are required to be single-valued, we must take

$$-\frac{\delta^2}{2\varepsilon} - \frac{2\nu+1}{2} = n, \qquad n = 0,1,2,3,... \tag{23}$$

which gives single-values wave functions. Using this requirement and further expanding Eq. (22) into series, we obtain

$$f(t) = N' \sum_{k=0}^{n} \frac{(-2\varepsilon)^k n!}{(n-k)! k!} (t+\varepsilon)^{-(2\nu+1+k)}, \tag{24}$$

where $N'$ is a constant. In terms of a simple extension of the inverse Laplace transformation [44, 45], we can immediately obtain

$$f(r) = N'' r^{2\nu} e^{-\varepsilon r} \sum_{k=0}^{n} \frac{(-1)^k n!}{(n-k)! k!} \frac{\Gamma(2\nu+1)}{\Gamma(2\nu+1+k)} (2\varepsilon r)^k, \tag{25}$$

and from $\varphi(r) = r^\alpha f(r)$, we then obtain

$$\varphi(r) = N''' r^\nu e^{-\varepsilon r} \sum_{k=0}^{n} \frac{(-1)^k n!}{(n-k)! k!} \frac{\Gamma(2\nu+1)}{\Gamma(2\nu+1+k)} (2\varepsilon r)^k, \tag{26}$$

where $N'''$ is a constant. On the other hand, the confluent hyper-geometric functions is defined as a series expansion [46]

$$_1F_1(-n, \gamma, s) = \sum_{j=0}^{n} \frac{(-1)^j n!}{(n-j)! j!} \frac{\Gamma(\gamma)}{\Gamma(\gamma+j)} s^j. \tag{27}$$

So, we find the upper-spinor component of wave function as

$$F_{n\kappa}(r) = N e^{-\varepsilon r} r^{\nu+1/2} L_n^{2\nu}(2\varepsilon r) = N \frac{\Gamma(n+2\nu+1)}{n!\Gamma(2\nu+1)} e^{-\varepsilon r} r^{\nu+1/2} {}_1F_1(-n, 2\nu+1, 2\varepsilon r), \qquad (28)$$

where $N$ is normalization constant. By using the normalization condition given as $\int_0^\infty |F_{n\kappa}(r)|^2 dr = 1$, and the relation between the Laguerre polynomials and confluent hyper-geometric functions as $L_n^p(x) = \frac{\Gamma(n+p+1)}{n!\Gamma(p+1)} {}_1F_1(-n, p+1, x)$ [45], the normalization constant in Eq. (28) is written

$$N = (2\varepsilon)^{\nu+1} \sqrt{\frac{n!}{\Gamma(n+2\nu+1)(2n+2\nu+1)}}, \qquad (29)$$

where we have used [45]

$$\int_0^\infty x^{q+1} e^{-x} L_n^q(x) L_{n'}^q(x) dx = (2n+q+1) \frac{\Gamma(n+q+1)}{n!} \delta_{nn'}. \qquad (30)$$

Inserting the parameters in Eqs.(16) and (17b) into Eq. (23), one obtain the energy eigenvalue of the radial part as follows

$$(m_0 c^2 - E_{n\kappa})(m_0 c^2 + E_{n\kappa} - C_s) = \frac{1}{4} \left[ \frac{q(m_0 c^2 + E_{n\kappa} - C_s) + b(C_s - 2m_0 c^2)}{n + \frac{1}{2} + \sqrt{\left(\kappa + T + \frac{1}{2}\right)^2 + b(b-q)}} \right]^2, \qquad (31)$$

$n = 0, 1, 2, ...$

which is identical to Ref. [18] when $2q \to q$ and the tensor interaction is removed, i.e., $T = 0$. For the constant mass case, we have

$$(m_0 c^2 - E_{n\kappa})(m_0 c^2 + E_{n\kappa} - C_s) = \frac{1}{4} \left[ \frac{q(m_0 c^2 + E_{n\kappa} - C_s)}{n + \kappa + T + 1} \right]^2, \quad n = 0, 1, 2, ..., \qquad (32)$$

In case when $C_s = 0$, $T = 0$, $\kappa = l$, $\Sigma(r) = V(r)$, $m_0 c^2 - E_{n\kappa} \to -E_{nl}$ and $m_0 c^2 + E_{n\kappa} \to m_0$, the above equation reduces to the non-relativistic energy limit (in units of $\hbar = 1$):

$$E_{nl} = -\frac{2m_0 q^2}{(n+l+1)^2}. \qquad (33)$$

The wave function (28) satisfies the boundary conditions, i.e., $F_{n\kappa}(r=0) = 0$ and $F_{n\kappa}(r \to \infty) \to 0$.

The lower spinor wave function can be obtained via

$$G_{n\kappa}(r) = \frac{1}{\left(m_0c^2 + \frac{\hbar cb}{r} + E_{n\kappa} - C_s\right)}$$

$$\times \left\{\left(-\varepsilon + \frac{v+1/2}{r} + \frac{\kappa+T}{r}\right)F_{n\kappa}(r) - 2\mathbf{N}\varepsilon e^{-\varepsilon r}r^{v+1/2}L_{n-1}^{2v+1}(2\varepsilon r)\right\}, \quad (27)$$

where we have used

$$\frac{d}{dx}L_n^\alpha(x) = -L_{n-1}^{(\alpha+1)}(x) = \frac{1}{x}\left[nL_n^\alpha(x) - (n+\alpha)L_{n-1}^\alpha(x)\right].$$

In tables 1 to 3 with $\hbar = c = 1$, we give some numerical results for the energy eigenvalues from energy formula (31).

## 2.2. p-spin symmetry case

To avoid repetition in the solution of Eq. (9b), the negative energy solution for p-spin symmetry can be obtained directly from those of the above positive energy solution for spin symmetry by using the parameter mapping [18]:

$$F_{n\kappa}(r) \leftrightarrow G_{n\kappa}(r); V(r) \to -V(r) \ (q \to -q); \kappa+1 \to \kappa; E_{n\kappa} \to -E_{n\kappa}; C_s \leftrightarrow -C_{ps}. \quad (28)$$

Following the previous results with the above transformations, we finally arrive at the energy equation as

$$(m_0c^2 + E_{n\kappa})(m_0c^2 - E_{n\kappa} + C_{ps}) = \frac{1}{4}\left[\frac{q(m_0c^2 - E_{n\kappa} + C_{ps}) + b(C_{ps} + 2m_0c^2)}{n + \sqrt{\left(\kappa+T-\frac{1}{2}\right)^2 + b(b+q)} + \frac{1}{2}}\right]^2, \quad (29)$$

$n = 0,1,2,...$

which is identical to Ref. [18] when $2q \to q$ and the tensor interaction is removed, i.e., $T = 0$. For the constant mass case, we have

$$(m_0c^2 + E_{n\kappa})(m_0c^2 - E_{n\kappa} + C_{ps}) = \frac{1}{4}\left[\frac{q(m_0c^2 - E_{n\kappa} + C_{ps})}{n+\kappa+T}\right]^2, \ n=0,1,2,..., \quad (30)$$

The lower-spinor component of wave function as

$$G_{n\kappa}(r) = \tilde{\mathbf{N}}e^{-\tilde{\varepsilon}r}r^{\tilde{v}+1/2}L_n^{2\tilde{v}}(2\tilde{\varepsilon}r) = \tilde{\mathbf{N}}\frac{\Gamma(n+2\tilde{v}+1)}{n!\Gamma(2\tilde{v}+1)}e^{-\tilde{\varepsilon}r}r^{\tilde{v}+1/2}{}_1F_1(-n,2\tilde{v}+1,2\tilde{\varepsilon}r), \quad (31)$$

where $\tilde{\varepsilon}^2 = \frac{1}{(\hbar c)^2}(m_0 c^2 + E_{n\kappa})(m_0 c^2 - E_{n\kappa} + C_{ps})$, $\tilde{v}^2 = \left(\kappa + T - \frac{1}{2}\right)^2 + b(b+q)$ and $\tilde{\mathbf{N}}$ is normalization constant

$$\tilde{\mathbf{N}} = (2\tilde{\varepsilon})^{\tilde{v}+1} \sqrt{\frac{n!}{\Gamma(n+2\tilde{v}+1)(2n+2\tilde{v}+1)}}, \tag{32}$$

The upper-spinor wave function can be obtained via

$$F_{n\kappa}(r) = \frac{1}{\left(m_0 c^2 + \frac{\hbar c b}{r} - E_{n\kappa} + C_{ps}\right)}$$

$$\times \left\{\left(-\tilde{\varepsilon} + \frac{\tilde{v}+1/2}{r} - \frac{\kappa+T}{r}\right)G_{n\kappa}(r) - 2\tilde{\mathbf{N}}\tilde{\varepsilon}e^{-\tilde{\varepsilon}r}r^{\tilde{v}+1/2}L_{n-1}^{2\tilde{v}+1}(2\tilde{\varepsilon}r)\right\}. \tag{33}$$

## 3. Numerical Results

In tables 1 to 3, we see that energies of bound states such as: $(np_{1/2}, np_{3/2}), (nd_{3/2}, nd_{5/2}), (nf_{5/2}, nf_{7/2}), (ng_{7/2}, ng_{9/2}), \ldots$ (where each pair is considered as a spin doublet) in the absence of the tensor potential are degenerate but in the presence of the tensor potential, the degeneracies are removed. Also, we investigate the effects of the $m_1$ and $C_s$ parameters on the bound states under the conditions of the spin symmetry limits for $T = 0$. The results are given in tables 1 to 3. It is readily seen that if $m_1$ and $C_s$ parameters increases, the value of the bound state energy eigenvalues of this potential increases for several states. It is shown that the energy eigenvalues decrease with decreasing $m_1$ and when increase with increasing the tensor strength $T$ for $\kappa > 0$ and $\kappa < 0$. The increase in the energies is slight when the strength of $T$ is large when its small when the strength $T$ is small. In Table 3, for constant values of $C_s$ and $T$, the energy increases when $m_1$ increasing. The decrease in the energy values is large without tensor interaction while small in presence of tensor interaction and then being large. Further, when $C_s$ increases, the energy increasing.

Finally, we plot the relativistic energy eigenvalues under spin and pspin symmetry limitations in figures 1 to 4. In fig. 2, we plot the energy eigenvalues of spin symmetry limit versus the perturbated mass $m_1$. It is seen that when $m_1$ increases, the energy increases too. In fig. 2, we have shown the variation of the energy as a function of $T$. We can see the degeneracy removes between spin doublets and also they become far from each other, when the parameter $T$ increases. In figs. 3 to 4, we plot the energy states of the pseudospin symmetry limit for different levels as functions of parameters $m_1$ and $T$, respectively. The variation of energy can also be seen from these figures.

## 4. Conclusion

In this paper, the relativistic equation for particles with spin 1/2 was solved exactly with both spatially-dependent mass and tensor interaction for attractive scalar and repulsive vector Coulomb potentials under the spin symmetry limit via the Laplace transformation method. Some numerical results are given for specific values of the model parameters. Effects of the tensor interaction on the bound states were presented that tensor interaction removes degeneracy between two states in spin doublets. We also investigated the effects of the spatially-dependent mass on the bound states under the conditions of the spin symmetry limits for $T = 0$.

**Table 1.** The bound state energy eigenvalues of the Coulomb potential under the P-spin symmetry limit for several values of $n$ and $\kappa$ for $m_0 = 5.0 fm^{-1}, \hbar = c = 1,, q = 1, T = (0,1,2,5),$ and $C_{ps} = -1.$

| $\tilde{l}$ | $n, \kappa < 0$ | $(l, j = l + 1/2)$ | $m_1$ | $E_{n,\kappa<0}$ | | | |
|---|---|---|---|---|---|---|---|
| | | | | $T=0$ | $T=1$ | $T=2$ | $T=5$ |
| 1 | 1, -2 | $1p_{3/2}$ | 0.0 | -4.86154 | -4.75676 | -4.47059 | -4.86154 |
| | | | 0.2 | -4.80508 | -4.66644 | -4.36211 | -4.80508 |
| | | | 0.4 | -4.74207 | -4.57172 | -4.26086 | -4.74207 |
| | | | 0.5 | -4.70859 | -4.52376 | -4.21286 | -4.70859 |
| 2 | 1, -3 | $1d_{5/2}$ | 0.0 | -4.91089 | -4.86154 | -4.75676 | -4.75676 |
| | | | 0.2 | -4.87335 | -4.80508 | -4.66644 | -4.66644 |
| | | | 0.4 | -4.83043 | -4.74207 | -4.57172 | -4.57172 |
| | | | 0.5 | -4.80716 | -4.70859 | -4.52376 | -4.52376 |
| 3 | 1, -4 | $1f_{7/2}$ | 0.0 | -4.93793 | -4.91089 | -4.86154 | -4.47059 |
| | | | 0.2 | -4.91138 | -4.87335 | -4.80508 | -4.36211 |
| | | | 0.4 | -4.88067 | -4.83043 | -4.74207 | -4.26086 |
| | | | 0.5 | -4.86385 | -4.80716 | -4.70859 | -4.21286 |
| 4 | 1, -5 | $1g_{9/2}$ | 0.0 | -4.95431 | -4.93793 | -4.91089 | -4.47059 |
| | | | 0.2 | -4.93461 | -4.91138 | -4.87335 | -4.36211 |
| | | | 0.4 | -4.91166 | -4.88067 | -4.83043 | -4.26086 |
| | | | 0.5 | -4.89903 | -4.86385 | -4.80716 | -4.21286 |
| 1 | 2, -2 | $2p_{3/2}$ | 0.0 | -4.91089 | -4.86154 | -4.75676 | -4.91089 |
| | | | 0.2 | -4.87402 | -4.80794 | -4.69083 | -4.87402 |
| | | | 0.4 | -4.83243 | -4.75015 | -4.62526 | -4.83243 |
| | | | 0.5 | -4.81014 | -4.72024 | -4.59287 | -4.81014 |
| 2 | 2, -3 | $2d_{5/2}$ | 0.0 | -4.93793 | -4.91089 | -4.86154 | -4.86154 |
| | | | 0.2 | -4.9116 | -4.87402 | -4.80794 | -4.80794 |
| | | | 0.4 | -4.88134 | -4.83243 | -4.75015 | -4.75015 |
| | | | 0.5 | -4.86487 | -4.81014 | -4.72024 | -4.72024 |
| 3 | 2, -4 | $2f_{7/2}$ | 0.0 | -4.95431 | -4.93793 | -4.91089 | -4.75676 |
| | | | 0.2 | -4.93469 | -4.9116 | -4.87402 | -4.69083 |
| | | | 0.4 | -4.91194 | -4.88134 | -4.83243 | -4.62526 |

| | | | 0.5 | -4.89945 | -4.86487 | -4.81014 | -4.59287 |
|---|---|---|---|---|---|---|---|
| 4 | 2, -5 | $2g_{9/2}$ | 0.0 | -4.96498 | -4.95431 | -4.93793 | -4.75676 |
| | | | 0.2 | -4.94984 | -4.93469 | -4.91160 | -4.69083 |
| | | | 0.4 | -4.93218 | -4.91194 | -4.88134 | -4.62526 |
| | | | 0.5 | -4.92244 | -4.89945 | -4.86487 | -4.59287 |

**Table 2.** The bound state energy eigenvalues of the Coulomb potential under the P-spin symmetry limit for several values of $n$ and $\kappa$ for $m_0 = 5.0 fm^{-1}$, $q = 1$, $\hbar = c = 1$, $T = (0,1,2,5)$, and $C_{ps} = -1$.

| $\tilde{l}$ | $n-1, \kappa > 0$ | $(l, j = l - 1/2)$ | $m_1$ | $E_{n-1,\kappa>0}$ | | | |
|---|---|---|---|---|---|---|---|
| | | | | $T=0$ | $T=1$ | $T=2$ | $T=5$ |
| 2 | 0, 2 | $0d_{3/2}$ | 0.0 | -4.75676 | -4.86154 | -4.91089 | -4.96498 |
| | | | 0.2 | -4.66644 | -4.80508 | -4.87335 | -4.94979 |
| | | | 0.4 | -4.57172 | -4.74207 | -4.83043 | -4.93205 |
| | | | 0.5 | -4.52376 | -4.70859 | -4.80716 | -4.92225 |
| 3 | 0, 3 | $0f_{5/2}$ | 0.0 | -4.86154 | -4.91089 | -4.93793 | -4.97231 |
| | | | 0.2 | -4.80508 | -4.87335 | -4.91138 | -4.96026 |
| | | | 0.4 | -4.74207 | -4.83043 | -4.88067 | -4.94614 |
| | | | 0.5 | -4.70859 | -4.80716 | -4.86385 | -4.93833 |
| 4 | 0, 4 | $0g_{7/2}$ | 0.0 | -4.91089 | -4.93793 | -4.95431 | -4.97756 |
| | | | 0.2 | -4.87335 | -4.91138 | -4.93461 | -4.96777 |
| | | | 0.4 | -4.83043 | -4.88067 | -4.91166 | -4.95628 |
| | | | 0.5 | -4.80716 | -4.86385 | -4.89903 | -4.94991 |
| 5 | 0, 5 | $0h_{9/2}$ | 0.0 | -4.93793 | -4.95431 | -4.96498 | -4.98144 |
| | | | 0.2 | -4.91138 | -4.93461 | -4.94979 | -4.97334 |
| | | | 0.4 | -4.88067 | -4.91166 | -4.93205 | -4.96381 |
| | | | 0.5 | -4.86385 | -4.89903 | -4.92225 | -4.95852 |
| 2 | 1, 2 | $1d_{3/2}$ | 0.0 | -4.86154 | -4.91089 | -4.93793 | -4.97231 |
| | | | 0.2 | -4.80794 | -4.87402 | -4.91160 | -4.96028 |
| | | | 0.4 | -4.75015 | -4.83243 | -4.88134 | -4.94621 |
| | | | 0.5 | -4.72024 | -4.81014 | -4.86487 | -4.93843 |
| 3 | 1, 3 | $1f_{5/2}$ | 0.0 | -4.91089 | -4.93793 | -4.95431 | -4.97756 |
| | | | 0.2 | -4.87402 | -4.91160 | -4.93469 | -4.96778 |
| | | | 0.4 | -4.83243 | -4.88134 | -4.91194 | -4.95632 |
| | | | 0.5 | -4.81014 | -4.86487 | -4.89945 | -4.94997 |
| 4 | 1, 4 | $1g_{7/2}$ | 0.0 | -4.93793 | -4.95431 | -4.96498 | -4.98144 |
| | | | 0.2 | -4.91160 | -4.93469 | -4.94984 | -4.97757 |

| | | | 0.4 | -4.88134 | -4.91194 | -4.93218 | -4.96383 |
| | | | 0.5 | -4.86487 | -4.89945 | -4.92244 | -4.95856 |
| 5 | 1, 5 | $1h_{9/2}$ | 0.0 | -4.95431 | -4.96498 | -4.97231 | -4.98440 |
| | | | 0.2 | -4.93469 | -4.94984 | -4.96028 | -4.97758 |
| | | | 0.4 | -4.91194 | -4.93218 | -4.94621 | -4.96957 |
| | | | 0.5 | -4.89945 | -4.92244 | -4.93843 | -4.96512 |

**Table 3.** The bound state energy eigenvalues of the Coulomb potential for several states under the p-spin symmetry limit with values of $C_{ps}$ with $m_0 = 5.0 fm^{-1}, q = 1, \hbar = c = 1, m_1 = (0, 0.5)$ and $T = (0, 5)$.

| $\tilde{l}$ | $n, \kappa$ | $(l, j = l+1/2)$ | $C_{ps}$ | $E_{n,\kappa<0}$ | | | |
| --- | --- | --- | --- | --- | --- | --- | --- |
| | | | | $T = 0$ $m_1 = 0$ | $T = 0$ $m_1 = 0.5$ | $T = 5$ $m_1 = 0$ | $T = 5$ $m_1 = 0.5$ |
| 1 | 1, 2 | $1d_{3/2}$ | -1.00 | -4.75676 | -4.52376 | -4.96498 | -4.92225 |
| | | | -1.25 | -4.76351 | -4.53699 | -4.96595 | -4.92441 |
| | | | -1.50 | -4.77027 | -4.55022 | -4.96693 | -4.92657 |
| | | | -1.75 | -4.77703 | -4.56345 | -4.96790 | -4.92873 |
| | | | -2.00 | -4.78378 | -4.57668 | -4.96887 | -4.93089 |
| | | | -2.25 | -4.79054 | -4.58991 | -4.96984 | -4.93305 |
| | | | -2.50 | -4.79730 | -4.60313 | -4.97082 | -4.93521 |
| | | | -2.75 | -4.80405 | -4.61636 | -4.97179 | -4.93736 |
| 3 | 0, -4 | $0f_{7/2}$ | -1.00 | -4.91089 | -4.80543 | -3.20000 | -2.90681 |
| | | | -1.25 | -4.91337 | -4.81083 | -3.25000 | -2.96496 |
| | | | -1.50 | -4.91584 | -4.81624 | -3.30000 | -3.0231 |
| | | | -1.75 | -4.91832 | -4.82164 | -3.35000 | -3.08124 |
| | | | -2.00 | -4.92079 | -4.82705 | -3.40000 | -3.13939 |
| | | | -2.25 | -4.92327 | -4.83245 | -3.45000 | -3.19753 |
| | | | -2.50 | -4.92574 | -4.83786 | -3.50000 | -3.25568 |
| | | | -2.75 | -4.92822 | -4.84326 | -3.55000 | -3.31382 |
| 1 | 1, -2 | $1p_{3/2}$ | -1.00 | -4.86154 | -4.70859 | -4.86154 | -4.70859 |
| | | | -1.25 | -4.86538 | -4.71669 | -4.86538 | -4.71669 |
| | | | -1.50 | -4.86923 | -4.72478 | -4.86923 | -4.72478 |
| | | | -1.75 | -4.87308 | -4.73288 | -4.87308 | -4.73288 |
| | | | -2.00 | -4.87692 | -4.74097 | -4.87692 | -4.74097 |
| | | | -2.25 | -4.88077 | -4.74907 | -4.88077 | -4.74907 |
| | | | -2.50 | -4.88462 | -4.75716 | -4.88462 | -4.75716 |

|   |      |            | -2.75 | -4.88846 | -4.76526 | -4.88846 | -4.76526 |
|---|------|------------|-------|----------|----------|----------|----------|
| 3 | 1, -4 | $1f_{7/2}$ | -1.00 | -4.93793 | -4.86385 | -4.47059 | -4.21286 |
|   |      |            | -1.25 | -4.93966 | -4.86764 | -4.48529 | -4.23472 |
|   |      |            | -1.50 | -4.94138 | -4.87142 | -4.50000 | -4.25659 |
|   |      |            | -1.75 | -4.9431  | -4.8752  | -4.51471 | -4.27845 |
|   |      |            | -2.00 | -4.94483 | -4.87898 | -4.52941 | -4.30032 |
|   |      |            | -2.25 | -4.94655 | -4.88276 | -4.54412 | -4.32218 |
|   |      |            | -2.50 | -4.94828 | -4.88654 | -4.55882 | -4.34405 |
|   |      |            | -2.75 | -4.95000 | -4.89033 | -4.57353 | -4.36591 |